%% file: main.tex
\documentclass[conference]{IEEEtran}
\IEEEoverridecommandlockouts

\usepackage{cite}
\usepackage{amsmath,amssymb,amsfonts}
\usepackage{graphicx}
\usepackage{textcomp}
\usepackage{booktabs}
\usepackage{array}
\usepackage{float}
\usepackage[hidelinks]{hyperref}
\usepackage{tikz}
\usepackage{siunitx}
\usepackage{xspace}
\usetikzlibrary{arrows.meta, calc}

\def\BibTeX{{\rm B\kern-.05em{\sc i\kern-.025em b}\kern-.08em
    T\kern-.1667em\lower.7ex\hbox{E}\kern-.125emX}}

\newcommand{\mctswivr}{\textit{MCTS-wIVR}\xspace}
\newcommand{\greedywivr}{\textit{Greedy-wIVR}\xspace}
\newcommand{\fixedpolicy}{\textit{Fixed}\xspace}
\newcommand{\mctsshort}{\textit{MCTS}\xspace}
\newcommand{\greedyshort}{\textit{Greedy}\xspace}

\begin{document}

\title{Real-Time Flight Test Maneuver Selection with Monte Carlo Tree Search}

\author{
\IEEEauthorblockN{Nicholas E. Bostock\textsuperscript{*\textdaggerdbl}, Helen Pruitt-Kennett\textsuperscript{*\textdagger\textdaggerdbl}, Marc R. Schlichting\textsuperscript{*}, and Mykel J. Kochenderfer\textsuperscript{*}}
\IEEEauthorblockA{\textsuperscript{*}Department of Aeronautics and Astronautics, Stanford University, Stanford, CA, 94305\\
\textsuperscript{\textdagger}Email: helenpk@stanford.edu\\
\textsuperscript{\textdaggerdbl}Equal contribution}
}

\maketitle

\begin{abstract}
Flight test is shifting toward a data-centric approach in which data contribute to model refinement, reducing reliance on pre-scripted test points. An open problem is how to sequence maneuvers within a sortie to maximize uncertainty reduction under resource limits. We present a real-time planning framework that combines a Gaussian Process (GP) belief model with Monte Carlo Tree Search (MCTS) to select pilot-actionable maneuvers under fuel constraints. Candidate maneuvers are scored using weighted integrated variance reduction (wIVR), and shallow lookahead is performed with a propagated per-evaluation-point variance state to account for downstream coverage redundancy and transition cost. The planner is evaluated in a closed, human-in-the-loop X-Plane simulation against greedy wIVR selection and a fixed test-card baseline. Sortie-summary statistics show significant directional differences, with \mctswivr achieving higher uncertainty reduction per unit fuel over both baselines. The results indicate that posterior-aware adaptive planning is a promising approach to increase efficiency of flight tests.
\end{abstract}


\section{Introduction}

Experimental flight test is a resource-intensive process central to every aircraft certification program. Despite advances in computational modeling and data analysis, test execution remains largely unchanged since the mid-20th century. Pilots are asked to hit discrete test points within narrow tolerances, and data collected outside those databands are discarded.

Harp et al. \cite{harp2025pigp} address this inefficiency through a data-centric architecture that replaces rigid test points with continuous, as-flown model refinement (Fig. \ref{fig:traditional_vs_pointless}). They use a Gaussian process~(GP) with a physics-informed mean function to model aerodynamic quantities directly from arbitrary flight data, eliminating databands while providing principled uncertainty quantification. This ``Pointless'' approach has been validated on the T-38C, recovering short-period dynamics to within the internal consistency of sixty years of historical data from a single sortie of low-frequency rollercoaster maneuvers.

While this architecture transforms \emph{what} is done with flight data, it leaves open the question of \emph{how} the data should be collected. Even with databands eliminated, a sortie remains finite: fuel burns, time passes, and the flight envelope imposes hard limits. A pilot who revisits well-characterized conditions wastes resources, while one who chases peak uncertainty without regard for transition costs may exhaust the sortie before resolving operationally critical regions. Currently, pilots and flight test engineers address this sequencing problem through pre-flight planning and in-flight judgment, with no principled, real-time guidance on which maneuver to fly next.

\input{figures/traditional_vs_pointless2}

Naive uncertainty-driven sampling strategies are insufficient in this setting because flight testing is fundamentally constrained and path-dependent. Maneuvers must satisfy aircraft dynamics, envelope limits, and fuel constraints, and the value of a candidate maneuver depends not only on its immediate information gain but also on the sequence of states required to reach and recover from it. Greedy policies that chase peak uncertainty can incur prohibitive transition costs, exhausting the sortie before critical regions are explored. Prior work on adaptive sampling and informative path planning addresses only part of this tension. Sampling-based motion planners such as RRT and Informed RRT primarily seek feasible or cost-improving trajectories in large state spaces, while informative path planning methods are typically developed for active mapping or sensing under uncertainty and travel budgets\cite{karaman2011rrt,gammell2014informed,popovic2020informative,akemoto2025informativepathplanningexplore,ott2024ipp}. As a result, these methods do not fully capture the coupled effects of vehicle feasibility, resource expenditure, and sequential information gain that define the flight test problem.

This paper introduces an online flight test planning framework that closes this gap. We integrate an existing GP belief model ~\cite{harp2024pointless,harp2025pigp} with Monte Carlo Tree Search~(MCTS) \cite{kochenderfer2022adm} to produce real-time, pilot-actionable maneuver recommendations throughout the sortie, making the following contributions: 
\begin{enumerate}    
    \item We formulate maneuver selection as a belief-state problem and approximate it as a belief-state  Markov decision process~(MDP) \cite{kochenderfer2022adm} during planning. We do this by holding the underlying GP posterior and hyperparameters fixed within the search tree, while propagating an approximate per-evaluation-point variance state through tree depth.

    \item We parameterize candidate actions as porpoising maneuvers and define a reward function that combines posterior uncertainty reduction with explicit penalties for fuel consumption.
    
    \item We enable multi-step lookahead that captures resource coupling---such as fuel consumption and Mach transition cost---across consecutive maneuvers, while also incorporating approximate progressive uncertainty reduction that penalizes redundant exploration of the evaluation envelope.
    
    \item We introduce a weighted integrated variance reduction (wIVR) acquisition function that scores candidate maneuvers by their expected posterior variance reduction over a fixed evaluation grid.
\end{enumerate}

Section~\ref{sec:method} describes the MCTS planning algorithm, the wIVR acquisition function, and the reward structure. Section \ref{sec:exp} details the closed, human-in-the-loop X-Plane simulation environment and the $C_m$ GP model used for evaluation. Section~\ref{sec:results} presents results comparing \mctswivr against greedy wIVR and a fixed test-card baseline, showing that shallow lookahead improves fuel normalized cumulative uncertainty reduction over greedy selection and the non-adaptive baseline.

\section{Background and Related Work}
\label{sec:background}

Traditional flight tests operate under a model-test-validate cycle in which data serve to spot-check pre-existing model predictions at discrete, precisely specified conditions~\cite{harp2024pointless}. Central to this paradigm is the test card: a fixed, predetermined sequence of discrete flight conditions specified by Mach number, altitude, and maneuver profile, determined by engineers prior to flight based on program requirements, envelope coverage targets, and historical precedent. Because the sequencing is decided entirely offline, the test card cannot adapt to what the data reveal about where uncertainty actually remains high, and sorties must be repeated when conditions are missed or data are deemed unusable.

Prior work inverted this relationship by making data the primary product of test~\cite{harp2024pointless,harp2025pigp}. Their architecture uses Gaussian Process Regression (GPR) with a physics-informed mean function derived from the Morelli generic aerodynamic model~\cite{grauer2015morelli}. An intentionally incorrect A-7E prior is refined using T-38C flight data, and the resulting GP hypersurface is differentiated with respect to state and control variables to obtain stability derivatives and short-period parameters as continuous functions of flight condition. The short-period mode describes the dominant longitudinal pitch response to control inputs, with natural frequency and damping ratio determined by aircraft geometry, mass properties, and stability derivatives (e.g., $M_{\alpha}$ and $M_Q$, the pitch moment sensitivities to angle of attack and pitch rate, respectively) \cite{phillips2009mechanics}. This approach demonstrates accurate short-period predictions from low-frequency maneuvers that lack the spectral content traditionally required by system identification methods, and it does so while correcting for control model mismatch between the A-7E and T-38C, validating the data assimilation process. 

GPs define a distribution over functions parameterized by a mean function and a covariance kernel, yielding posterior predictive means and variances at arbitrary query points~\cite{rasmussen2006gp,kochenderfer2019}. Although the GP architecture provides a principled mechanism for building aerodynamic models from arbitrary flight data, it does not prescribe which data to collect. 
A GP belief model trained on as-flown data provides calibrated confidence estimates across the flight envelope, enabling principled identification of high uncertainty regions where additional data would be most informative. Shown in this work, this enables active learning strategies in which future data collection is guided by epistemic uncertainty rather than being passively recorded. In particular, the posterior covariance of the GP provides a principled measure of epistemic uncertainty, which can be used to evaluate the expected information gain of candidate maneuvers.

\subsection{Monte Carlo Tree Search for Sequential Decision-Making}
MCTS is a simulation-based planning algorithm for sequential decision-making in large or continuous action spaces~\cite{kochenderfer2022adm,browne2012mcts}. The algorithm constructs an asymmetric search tree by iterating four phases: selection, expansion, simulation~(rollout), and backpropagation. Node selection uses the Upper Confidence Bound for Trees (UCT) criterion, which balances exploitation of high-value actions against exploration of
less-visited branches~\cite{browne2012mcts}. Because MCTS operates on a generative forward model rather than an explicit value function, it is well suited for problems where belief dynamics are approximated rather than computed exactly~\cite{kochenderfer2022adm}.

MCTS has been applied to robotic path planning~\cite{karaman2011rrt}, adaptive experiment design~\cite{vanlier2012bayesian}, and trajectory optimization with obstacle avoidance~\cite{webb2013kinodynamic}. The multi-step lookahead capability of MCTS is particularly valuable when the optimal sequence of decisions differs from the sequence produced by repeatedly taking the locally greedy action, which is especially consequential in resource-constrained settings where early high-cost actions can preclude later high-value ones. Within the aerospace domain, MCTS has been applied to falsification of longitudinal hybrid flight control laws~\cite{delmas2019mcts} and real-time planning for continuous dynamical systems including spacecraft and ground vehicles using receding-horizon tree search~\cite{riviere2024montecarlo}. 

\subsection{Active Learning and Belief-Space Planning}
Active learning formalizes the adaptive selection of query points to reduce model uncertainty, typically through information-theoretic criteria such as entropy reduction or mutual information~\cite{settles2009active,paninski2005infodesign}. Standard greedy active learning selects the single point of maximum uncertainty at each step, ignoring transition costs and the effect of current choices on future opportunities. This myopia is particularly costly in flight test, where transitioning between flight conditions consumes fuel, and where the order in which conditions are sampled affects what remains feasible later in the sortie. A fixed test card avoids this myopia through pre-flight planning, but cannot respond to what the data reveal in real time. Greedy active learning responds to the data but ignores the multi-step consequences of each choice. The framework presented here aims to address both limitations simultaneously.

In contrast with pointwise formulations, this work evaluates the expected information gain of entire maneuver trajectories rather than individual query points. Each candidate maneuver is assessed by its aggregate effect on posterior uncertainty across the evaluation envelope, which accounts for the structured information content of continuous trajectories that pointwise strategies cannot capture.

Belief-space planning addresses the myopia of greedy strategies by explicitly modeling the evolution of the posterior belief over a planning horizon~\cite{silver2010pomcp,ross2008online}. MCTS is a natural solver for this formulation: it does not require an analytic expression for the value function and can accommodate the non-stationary belief dynamics that result from Bayesian updating.

The combination of GP belief models with MCTS-based planning has been
explored for underwater terrain mapping~\cite{guo2022auv} and planetary
surface exploration using GP-based informative path
planning~\cite{akemoto2025informativepathplanningexplore}. Active learning methods have also
been applied to the identification of nonlinear dynamical systems,
demonstrating that adaptive input design can achieve substantially
greater sample efficiency than passive
excitation. The presented work differs from prior GP-based informative planning approaches by evaluating information gain over parameterized flight-test maneuvers, incorporating operational constraints such as fuel consumption and Mach repositioning cost, and applying MCTS to select sequences of maneuvers under these constraints in real time.

\section{Approach}
\label{sec:method}

We fit a GP model which learns the mapping from flight condition $\mathbf{x}$ to an aerodynamic quantity $f(\mathbf{x})$. The GP posterior at a query point $\mathbf{x}'$ is given
by~\cite{rasmussen2006gp}
\begin{equation}
\mu(\mathbf{x}') = \mathbf{m}(\mathbf{x}') + \mathbf{K}(\mathbf{x}', X)
\left[\mathbf{K}(X,X) + \sigma_n^2 \mathbf{I}\right]^{-1}
(\mathbf{y} - \mathbf{m}(X)),
\end{equation}
\begin{equation}
\sigma^2(\mathbf{x}') = \mathbf{K}(\mathbf{x}', \mathbf{x}') -
\mathbf{K}(\mathbf{x}', X)
\left[\mathbf{K}(X,X) + \sigma_n^2 \mathbf{I}\right]^{-1}
\mathbf{K}(X, \mathbf{x}').
\end{equation}
Here, $X$ denotes the training data, $\mathbf{y}$ are the
observed values, $\mathbf{m}(\cdot)$ is the prior mean function,
$\mathbf{K}(\cdot,\cdot)$ is the covariance kernel, and $\sigma_n^2$ is
the observation noise variance. Rather than using a zero-mean prior, we incorporate a physics-informed mean function, encoding known structure and improving data efficiency \cite{harp2025pigp}.

We apply a standard MCTS objective approximation that maximizes cumulative discounted reward over a sequence of actions. Action selection during tree expansion balances exploitation of high-value actions and exploration of uncertain regions via the UCB1 rule. The resulting objective is
\begin{equation}
\label{eq:objective}
a^{*} = \arg\max_{a \in \mathcal{A}}
\; \mathbb{E}\!\left[\sum_{t=0}^{H} \gamma^t R(s_t, a_t)\right],
\end{equation}
where $\gamma$ is the discount factor and $H$ is the horizon. Action selection follows the UCB1 exploration rule
\begin{equation}
\label{eq:uct_method}
a^{*} =
\arg\max_{a}\left\{
Q(s,a) + c \sqrt{\frac{\ln N(s)}{N(s,a)}}
\right\},
\end{equation}
where $Q(s,a)$ is the estimated action value, $N(s)$ is the total visit count for state $s$, $N(s,a)$ is the visit count for the state-action pair, and $c = \sqrt{2}$ is the exploration constant~\cite{kochenderfer2022adm}.

\subsection{Monte Carlo Tree Search Planner}
The underlying problem is naturally partially observable because the true function is unknown and is inferred from noisy flight data through a GP posterior. We approximate this problem as a belief-state MDP during search by holding the underlying GP posterior and hyperparameters fixed within the MCTS tree, while propagating an approximate per-evaluation-point variance state through tree depth. Under this approximation, multi-step lookahead captures coupling in fuel consumption, Mach repositioning, and approximate progressive uncertainty reduction, with diminishing returns enforced when consecutive maneuvers cover overlapping regions of the evaluation envelope.

The one-step reward for executing action $a$ from belief-state $s$ is
\begin{equation}
\label{eq:reward_formulation}
R(s,a) =
\frac{J_{\mathrm{acq}}(s,a)}{\tilde{J}_{\mathrm{acq}}(b)}
- \lambda_{\mathrm{fuel}} J_{\mathrm{fuel}}(s,a)
- \lambda_{\mathrm{trans}} J_{\mathrm{trans}}(s,a),
\end{equation}
where $J_{\mathrm{acq}}$ is the acquisition score of the maneuver under the current GP posterior, normalized by $\tilde{J}_{\mathrm{acq}}$, the median acquisition value over all candidate actions; $J_{\mathrm{fuel}}$ is the normalized maneuver fuel cost; and $J_{\mathrm{trans}}$ is the normalized Mach repositioning cost relative to the current Mach number. The weights $\lambda_{\mathrm{fuel}}$ and $\lambda_{\mathrm{trans}}$ control the trade-off between information gain and resource usage.

Within the tree search, the underlying GP posterior and hyperparameters are held fixed across all tree nodes. Rather than re-conditioning the GP at every node, the planner precomputes a per-evaluation-point variance-reduction vector $\Delta\boldsymbol{\sigma}^2_a \in \mathbb{R}^{N_{\mathrm{eval}}}$ for each candidate action at the root via hypothetical conditioning on the action's visited states. A variance state $\boldsymbol{\sigma}^2_s$ is then propagated through tree depth: when action $a$ is applied at state $s$, the in-tree acquisition value is evaluated against the node's current variance state,
\begin{equation}
J_{\mathrm{acq}}(s, a) = \sum_{i=1}^{N_{\mathrm{eval}}} w_i \min\!\left(\Delta\sigma^2_{a,i},\; \sigma^2_{s,i}\right),
\end{equation}
and the child state carries the reduced variance $\sigma^2_{s',i} = \max(\sigma^2_{s,i} - \Delta\sigma^2_{a,i},\, 0)$. The $\min(\cdot)$ clamp enforces that an action cannot reduce variance below what the node already holds, which produces undesirable behavior in-tree: actions covering regions that earlier tree steps have already reduced score lower than they would at the root. This design captures approximate progressive uncertainty reduction across the planning horizon at $O(N_{\mathrm{eval}})$ per tree node, without the $O(n^3)$ cost that full GP re-conditioning would incur.
The normalized fuel cost is
\begin{equation}
J_{\mathrm{fuel}}(s,a) = \frac{f_{\mathrm{base}}(a) + f_{\mathrm{trans}}(a)}{f_{\mathrm{rem}}(s)},
\end{equation}
where $f_{\mathrm{rem}}$ is  the remaining fuel at belief-state $s$ while $f_{\mathrm{base}}$ and $f_{\mathrm{trans}}$ estimate fuel consumed during action $a$. The base fuel cost is computed by integrating instantaneous fuel flow over the maneuver duration,
\begin{equation}
f_{\text{base}}(a) = \dot{m}_{\mathrm{fuel}} \, t_{\mathrm{maneuver}},
\end{equation}
where the fuel flow rate $\dot{m}_{\mathrm{fuel}}$ is obtained from a log-linear least-squares fit to baseline maneuver data (see Appendix~\ref{app:fuel}). 

We use a proxy for fuel cost associated with transitioning between airspeeds. It is modeled as

\begin{equation}
\begin{aligned}
f_{\mathrm{trans}}(a) =\;& \max(\Delta U_0, 0)\, k_{\mathrm{accel}}\, \dot{m}_{\mathrm{fuel}} \\
&+ \left|\min(\Delta U_0, 0)\right|\, k_{\mathrm{decel}}\, \dot{m}_{\mathrm{fuel}},
\end{aligned}
\end{equation}
where $k_{\mathrm{accel}}$ and $k_{\mathrm{decel}}$ are acceleration and deceleration fuel-per-knot ratios, respectively.

The Mach repositioning cost is
\begin{equation}
J_{\mathrm{trans}}(s,a) = \frac{|M(a) - M_{\mathrm{cur}}(b)|}{M_{\max} - M_{\min}},
\end{equation} 
where $M_{\mathrm{cur}}$ is the current belief-state's Mach number, and $M_{\max}$ and $M_{\min}$ are the flight envelope maximum and minimum Mach numbers. These terms discourage excessively long or energetically costly trajectories within the MCTS planning horizon.

Normalizing $J_{\mathrm{acq}}$ by $\tilde{J}_{\mathrm{acq}}$ renders the information term scale-invariant so that the resource-penalty weights $\lambda_{\mathrm{fuel}} = 0.02$ and $\lambda_{\mathrm{trans}} = 0.005$ retain consistent interpretation across planning iterations as the absolute magnitude of achievable variance reduction changes with accumulated data. Because acquisition values are recomputed against a variance state that is propagated through tree depth, the multi-step lookahead captures both resource-coupling effects---fuel consumption and Mach repositioning---and information gain, penalizing action sequences that visit redundant regions of the evaluation envelope.

\section{Experimental Setup}
\label{sec:exp}
We assess the effectiveness of the proposed framework using the X-Plane 12 simulation environment. X-Plane is widely used for research and training, and was used here to evaluate the maneuver selection planner in a dynamic setting. This section details the experimental conditions we used to refine a moment coefficient GP model, including the simulated sortie execution, the operational loop, the specific GP model, and the MCTS planner.

\subsection{Sortie Execution}
\label{subsec:sortie}

\begin{figure}
    \centering
    \includegraphics[width=.8\linewidth]{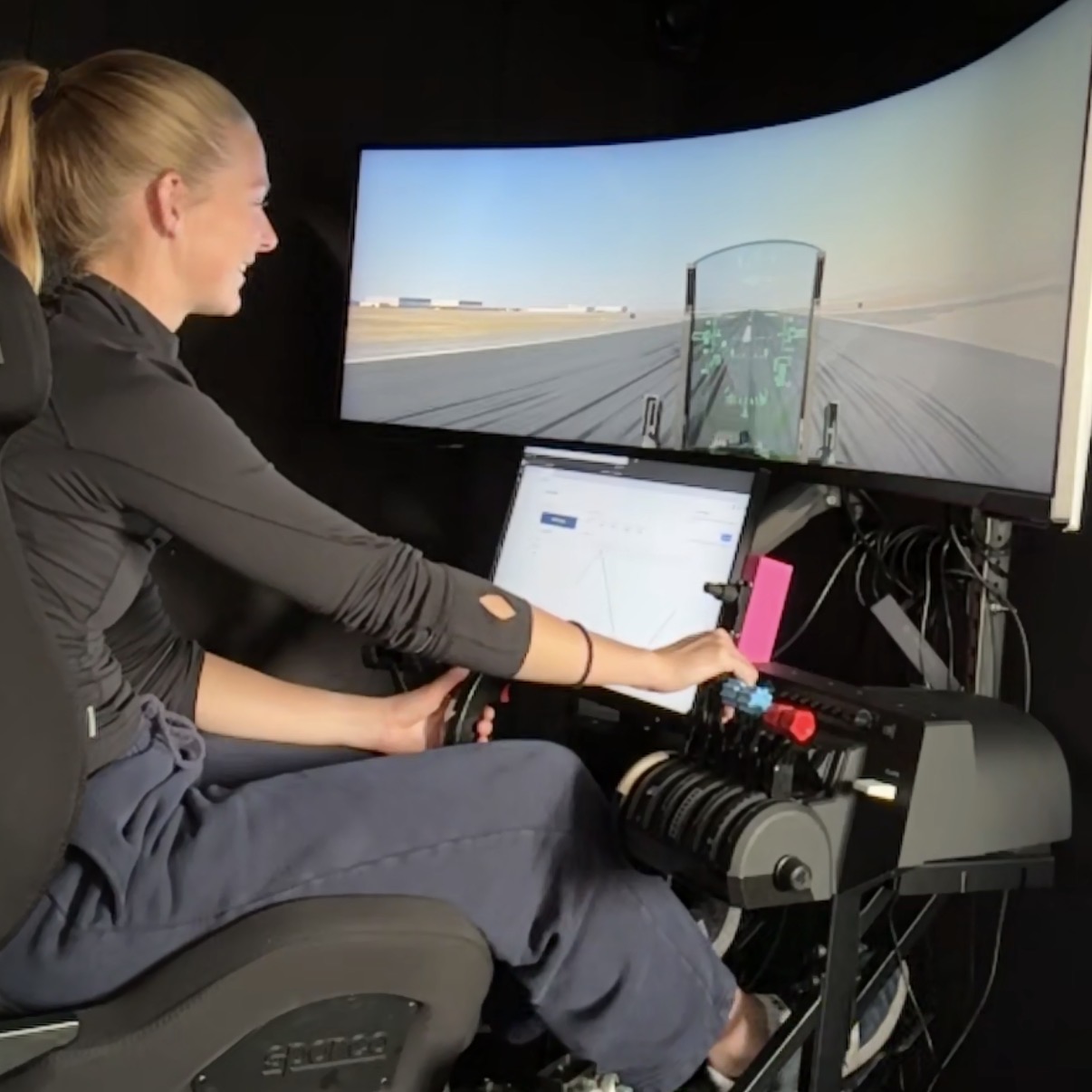}
    \caption{Sortie execution in simulator using X-Plane 12 and F-16C.}
    \label{fig:sim}
\end{figure}

The aerodynamic GP model focused exclusively on the pitch moment coefficient; accordingly, the action space was restricted to maneuvers that excite only longitudinal states. Each maneuver consisted of a single porpoise, in which the pilot smoothly increased and decreased the load factor to the prescribed maximum and minimum G-force while maintaining airspeed. All maneuvers were book-ended with 1.5 seconds of steady, wings-level flight and started at 15,000 ft. The samples were flown on X-Plane 12 using the F-16C and the data was measured at 5 Hz, shown in Fig. \ref{fig:sim}. A single CFI pilot flew each sortie.

All sorties in the experiment were initialized from a fixed, shared set of three baseline maneuvers common to all policies and runs, and they were used to fit the initial GP across the subsonic regime. These porpoises had G-profiles with $g_\text{peak}$, $g_\text{min}$, $\dot{g}{\uparrow}$, and $\dot{g}{\downarrow}$ of 2.0 g, $-1.0$ g, $0.5$ g/s, and $-0.5$ g/s, respectively, and Mach numbers of 0.5, 0.7, and 0.9. Following this shared initialization, actions were sequentially selected to minimize model uncertainty.

The performance of the proposed approach was compared against two baseline policies: greedy wIVR and a fixed test-card. Eight sorties were flown per policy, each consisting of ten maneuvers---three baseline initialization maneuvers plus seven planner selected follow-up maneuvers. The fixed test-card's deterministic schedule yields identical command sequences across runs; variation between \fixedpolicy sorties therefore reflects only X-Plane dynamics and pilot execution noise. Table~\ref{t:test_card_actions} details the seven fixed test-card maneuvers, designed to progressively increase structural loading through the sortie.

\begin{table}
\centering
\caption{Fixed test-card action set}
\label{t:test_card_actions}
\begin{tabular}{@{}l
                S[table-format=1.1]
                S[table-format=1.1]
                S[table-format=2.1]
                S[table-format=1.1]
                S[table-format=2.1]@{}}
\toprule
Action & {$M$} & {$\dot{g}_{\uparrow}$} & {$\dot{g}_{\downarrow}$} & {$g_{\text{peak}}$} & {$g_{\text{min}}$} \\
\midrule
$a_1$ & 0.6 & 0.5 & -0.5 & 2.0 & -0.5 \\
$a_2$ & 0.9 & 0.5 & -0.5 & 2.0 & -0.5 \\
$a_3$ & 1.0 & 0.5 & -0.5 & 2.0 & -0.5 \\
$a_4$ & 0.7 & 1.5 & -1.5 & 3.0 & -0.5 \\
$a_5$ & 0.9 & 1.5 & -1.5 & 3.0 & -0.5 \\
$a_6$ & 1.0 & 1.5 & -1.5 & 3.0 & -0.5 \\
$a_7$ & 0.5 & 1.5 & -1.5 & 4.0 & -0.5 \\
\bottomrule
\end{tabular}
\end{table}

\subsection{Pilot Interface and Operational Loop}

The system is implemented as a web application connected to the simulator via UDP. The planner is executed offboard in simulation and provides maneuver recommendations to the pilot via a ground-based interface; no onboard flight-critical computation is assumed. The pilot and test director interact through a single-page interface that progresses through two phases.

Baseline phase: Three fixed maneuvers are flown first, consisting of a 2g pull at Mach 0.5, 0.7, and 0.9, to establish the initial GP posterior. For each maneuver, the interface presents a maneuver card specifying the commanded parameters (Mach, pull rate, peak load factor, unload rate, and minimum load), along with a synthesized normal load factor profile with limit overlays. The pilot initiates and terminates recording before and after executing each maneuver. Once all baseline maneuvers are completed, the operator triggers GP hyperparameter optimization on the backend.

\begin{figure}[H]
    \centering
    \includegraphics[width=\linewidth]{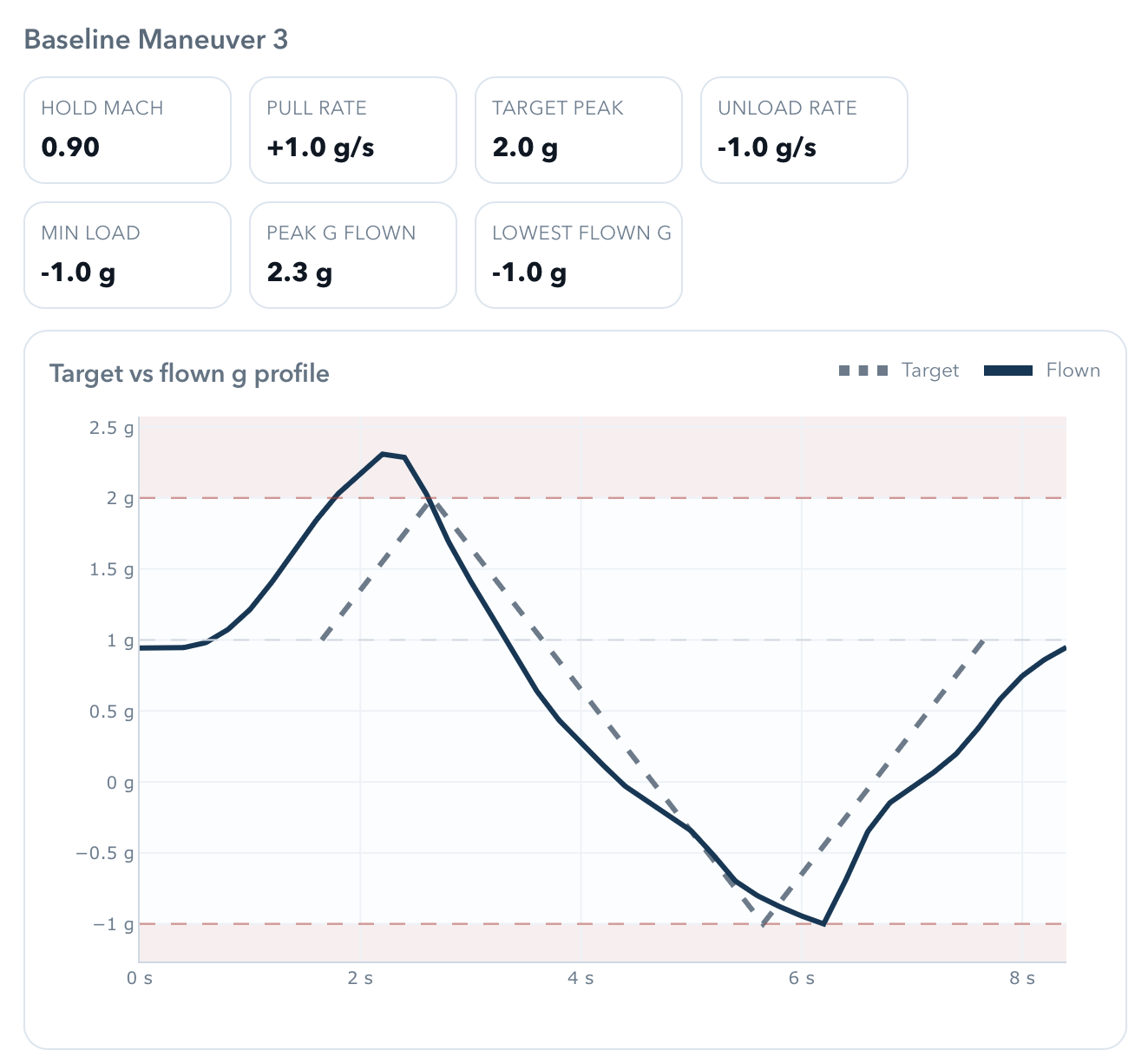}
    \caption{User interface screenshot illustrating a Target vs As-Flown G-profile.}
    \label{fig:ui}
\end{figure}

Adaptive test-point loop: After initialization, the system enters a closed-loop process repeated for each test point:
\begin{itemize}
    \item \textbf{Plan.} The backend executes the selected planner (\mctsshort, \greedyshort, or \fixedpolicy). For \mctsshort, a progress indicator displays completed simulations relative to the computational budget.
    \item \textbf{Present.} The interface displays a recommended maneuver card with commanded parameters and corresponding load factor profile (Fig. \ref{fig:ui}).
    \item \textbf{Fly and record.} The pilot executes the maneuver while the system streams simulator data at 5 Hz, including angle of attack, Mach number, body rates, control inputs, normal load factor, and fuel flow.
    \item \textbf{Update.} The recorded data are processed to update the GP posterior. A debrief visualization is presented, comparing commanded and flown profiles with tracking error feedback.
    \item \textbf{Repeat.} The planner re-evaluates using the updated posterior, and the loop continues until termination criteria are met.
\end{itemize}

The maneuver parameterization fully specifies a flyable trajectory, which is precomputed into a state sequence prior to execution. This ensures that the pilot receives concrete, operationally meaningful instructions rather than abstract optimization outputs.

\subsection{GP Model: $C_m(\mathbf{x})$ }
We represented flight conditions with an eight-dimensional state vector
\begin{equation}
\label{eq:state_vector}
\mathbf{x} =
\begin{bmatrix}
M & \rho & \bar{q} & P & Q & R & \alpha & \delta_e
\end{bmatrix}^\top,
\end{equation}
where $M$ is Mach number, $\rho$ air density, $\bar{q}$ dynamic pressure, $P$, $Q$, $R$ body-axis angular rates, $\alpha$ angle of attack, and $\delta_e$ stabilator deflection.
We followed the GP framework outlined in Section \ref{sec:method} and used an existing mean function \cite{harp2024pointless}. The physics-informed
Morelli mean function~\cite{grauer2015morelli} was used to incorporate
known aerodynamic structure:
\begin{equation}
\label{eq:morelli}
\begin{split}
\mathbf{m}(\mathbf{x}) =\ & \theta_{29} + \theta_{30}\alpha +
\theta_{31}\tilde{q} + \theta_{32}\delta_e \\
& + \theta_{33}\tilde{q}\alpha + \theta_{34}\tilde{q}\alpha^2 +
\theta_{35}\delta_e\alpha^2 \\
& + \theta_{36}\tilde{q}\alpha^3 + \theta_{37}\delta_e\alpha^3 +
\theta_{38}\alpha^4,
\end{split}
\end{equation}
where $\tilde{q} = q\bar{c}/(2U_0)$ is the nondimensional pitch rate for $q$, $\bar{c}$ is the mean aerodynamic chord, $U_0$ is the free stream velocity, and $\{\theta_i\}$ are the A-7E coefficients. This mean function encodes
nonlinear aerodynamic couplings, leaving the GP to model residuals due
to configuration differences and unmodeled effects.

We used the Mat\'ern $\nu$ = 5/2 covariance kernel function with automatic relevance determination (ARD):
\begin{equation}
k(\mathbf{x}, \mathbf{x}') =
\sigma^2
\left(1 + \sqrt{5}\, r + \frac{5}{3} r^2 \right)
\exp(-\sqrt{5}\, r)
\end{equation}
where
\begin{equation}
r = \sqrt{
\sum_{i=1}^8
\left(\frac{x_i - x_i'}{\ell_i}\right)^2
}.
\end{equation}

We scaled all inputs to $[0,1]$ using min-max normalization. The Nelder--Mead algorithm \cite{Nelder1965} optimizes the length scales $\ell_i$, while the observation noise variance is held fixed at $\sigma_n^2 = 10^{-4}$ to decouple noise estimation from length-scale optimization.

\subsection{MCTS Planner}
Each action $a \in \mathcal{A}$ defines a porpoising maneuver parameterized by
\begin{equation}
a = \left[M,\; \dot{g}_{\uparrow},\; \dot{g}_{\downarrow},\;
          g_{\text{peak}},\; g_{\text{min}}\right],
\end{equation}
where $M$ denotes the target Mach number, $\dot{g}_{\uparrow}$ and $\dot{g}_{\downarrow}$ represent the pull-up and push-over load factor rates (g/s), respectively, and $g_{\text{peak}}$ and $g_{\text{min}}$ define the maximum and minimum load factors (g). Every action additionally carries a simulated maneuver trajectory in the GP input space, represented as a sequence of visited flight states, along with a base fuel cost and a transition cost. 

MCTS rollouts proceeded to a fixed depth of $d = 2$ with a discount factor of $\gamma = 0.95$, and leaf nodes at the depth horizon return a terminal value of zero. Each planning call executed $m = 5|\mathcal{A}| = 680$ simulations before selecting the root action with highest estimated action value. Within the cost structures, $M_{\max}$ and $M_{\min}$ were 1.8 and 0.4 while $k_{\mathrm{accel}}$ and $k_{\mathrm{decel}}$ were 0.66 and 0.49, respectively.  

\begin{table}[h]
\centering
\caption{Maneuver parameter space}
\label{t:actions}
\begin{tabular}{@{}ll@{}}
\toprule
Parameter & Discrete set \\
\midrule
$M$ & $\{0.5, 0.6, \ldots, 1.0\}$ \\
$g_{\text{peak}}$ & $\{2, 3, 4\}\ \mathrm{g}$ \\
$g_{\text{min}}$ & $\{-0.5, -1.0\}\ \mathrm{g}$ \\
$\dot{g}_{\uparrow}$ & $\{0.5, 1.5\}\ \mathrm{g/s}$ \\
$\dot{g}_{\downarrow}$ & $\{-0.5, -1.5\}\ \mathrm{g/s}$ \\
\bottomrule
\end{tabular}
\end{table}

The action space consists of 144 distinct Mach, maximum/minimum load factor, and pull-up/push-over rate action combinations ($6 \times 3 \times 2 \times 2 \times 2$). Of the 144 actions, 8 were pruned since 0.5 $M$ was not maintainable at $g_{\text{peak}}$ = 4, leaving 136 candidate actions. Table \ref{t:actions} summarizes the maneuver parameter space and Fig. \ref{fig:ui} illustrates a user interface highlighting a plotted maneuver target and as-flown g-profile.

Each action specifies a load factor profile $G(t)$ and a Mach number. Within the tree search, the set of visited flight states along each candidate maneuver is used to evaluate the wIVR acquisition through hypothetical GP conditioning.

To maintain computational tractability, several simplifying assumptions are introduced to reduce the dimensionality of the problem. Density altitude was held constant, and roll and yaw rates were assumed to be zero. Under a standard-day atmosphere with fixed density altitude, dynamic pressure is a deterministic function of Mach number alone. The stabilator trim required for steady, wings-level flight was then computed as a function of dynamic pressure for each candidate condition.

Under these assumptions, the action space was reduced to the variables $\{M, Q, \alpha\}$. With Mach number specified by the action and held constant during the maneuver, the pitch rate and angle of attack are deduced from the prescribed G-force profile as shown in Fig. \ref{fig:ui}.

The pitch rate is approximated as
\begin{equation}
Q(t) = \frac{(G(t) - 1)\, g}{M(t) a_\infty},
\end{equation}
where $a_\infty$ is the freestream speed of sound and $g$ is gravitational acceleration. This relation follows from the approximation $a_n \approx V Q$ for small-angle, coordinated flight, with $(G-1)g$ representing the maneuver-induced normal acceleration $a_n$.
Angle of attack is then obtained from subsonic lift models~\cite{phillips2009mechanics}. For subsonic conditions ($M < 1$),
\begin{equation}
\alpha(t) =
\frac{C_L(G(t)) - C_{L0}}{a_0 \sqrt{1 - M^2}},
\end{equation}
where $C_{L0}$ is the zero-angle-of-attack lift coefficient, $C_{L}(G(t))$ denotes the lift coefficient at $G(t)$ and $a_0$ is an empirical regression coefficient obtained from the baseline aerodynamic fit in the subsonic regime. Data from the three subsonic baseline maneuvers, introduced in Section \ref{subsec:sortie}, inform these coefficients through inversion of a least-squares lift model, as described in Appendix~\ref{app:lift}. 

The evaluation grid was discretized as a structured Cartesian lattice over three dimensions: Mach, angle of attack, and pitch rate. Specifically, Mach was sampled from $0.5$ to $1.0$ in $0.1$ increments (6 points), $\alpha$ from $-12^\circ$ to $+12^\circ$ in $2^\circ$ steps (13 points), and $Q$ from $-10^\circ/\mathrm{s}$ to $+10^\circ/\mathrm{s}$ in $5^\circ/\mathrm{s}$ steps (5 points), giving $6 \times 13 \times 5 = 390$ evaluation states. For each Mach slice, dynamic pressure was computed from the reference density and sound speed, the stabilator trim was set from that dynamic pressure, and the remaining state channels ($P$ and $R$) were fixed at zero, producing each 8-state grid vector. These evaluation points approximate the variance reduction across the GP model within the tree search.

\section{Results}
\label{sec:results}

This section evaluates whether adaptive maneuver selection reduces GP posterior uncertainty more effectively than myopic and a non-adaptive baseline under a fixed sortie budget. The primary outcomes are sortie-level uncertainty reduction, fuel-normalized efficiency, and final posterior uncertainty after the allocated maneuver sequence.

All policies begin from a common initialization of three baseline maneuvers and then select up to seven additional maneuvers. Three policies are compared: \mctswivr ($n = 8$), \greedywivr ($n = 8$), and a \fixedpolicy test-card ($n = 8$). The \mctsshort-versus-\greedyshort comparison isolates the effect of lookahead under a shared wIVR acquisition rule. \greedyshort selects maneuvers from one-step acquisition values, whereas \mctsshort evaluates multi-step sequences ($d = 2$) with the resource-aware objective and propagated variance state (Section~\ref{sec:method}), explicitly accounting for downstream fuel and transition costs.

\sisetup{
    round-mode = figures,
    round-precision = 3,
    separate-uncertainty = true,
    table-align-uncertainty = true
}

\begin{table}[h]
\centering
\caption{Sortie uncertainty reduction and fuel consumption with standard deviations}
\label{t:sortie_stats}
\setlength{\tabcolsep}{3pt}

\begin{tabular}{
    @{}
    l
    S[table-format=1.3(2)]
    S[table-format=1.2(2)]
    S[table-format=4.0(2)]
    @{}
}
\toprule
Method & {$\Delta U_{\text{tot}}$} & {$\Delta U/\text{fuel}~(\times 10^{-4})$} & {Fuel (lb)} \\
\midrule
\mctswivr & 0.342(26) & 4.54(9)   & 753(64) \\
\greedywivr & 0.317(25) & 3.41(17)  & 929(76) \\
\fixedpolicy & 0.358(51) & 2.15(31)  & 1667(55) \\
\bottomrule
\end{tabular}
\end{table}

Table \ref{t:sortie_stats} summarizes sortie-level uncertainty reduction and fuel usage over eight runs per policy. In this sample, the \fixedpolicy baseline achieves the highest mean total uncertainty reduction ($\Delta U_{\text{tot}}$), while both adaptive policies use substantially less fuel and therefore attain higher fuel-normalized reduction than \fixedpolicy. Specifically, \mctswivr yields the largest mean $\Delta U/\text{fuel}$ among the three methods.

One-sided Welch tests using the means and standard deviations in Table~\ref{t:sortie_stats} (with $n=8$ per policy) show statistically significant directional effects at $\alpha=0.05$ for fuel-normalized reduction, where \mctswivr exceeds \greedywivr and \fixedpolicy, and \greedywivr exceeds \fixedpolicy. For total uncertainty reduction, \fixedpolicy exceeds both adaptive policies, and \mctswivr significantly outperforms \greedywivr ($p = 0.034$).

\begin{figure}[H]
    \centering
    \includegraphics[width=\linewidth]{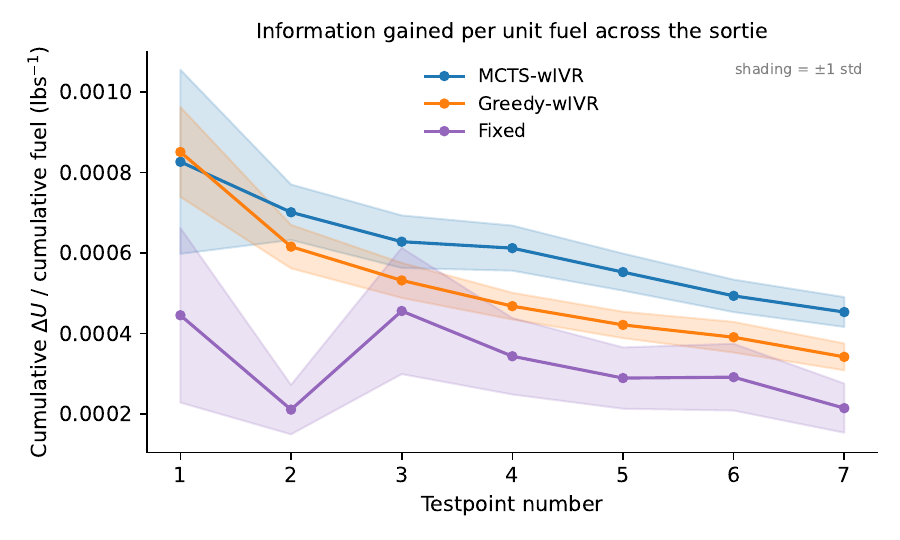}
    \caption{Cumulative uncertainty reduction $\sum_{i=1}^{k}\Delta U_i$ versus maneuver index. Shaded bands denote $\pm 1$ standard deviation across trials.}
    \label{fig:cum_delta_u_vs_pt}
\end{figure}

Fig. \ref{fig:cum_delta_u_vs_pt} shows the mean cumulative uncertainty reduction normalized by consumed fuel over the sortie. Adaptive methods exhibit a monotonic decrease across the seven post-initialization maneuvers while \fixedpolicy generally decreases. \mctswivr and \greedywivr remain close in early maneuvers, but \mctswivr more efficiently reduces uncertainty on a fuel cost basis.

In the current experimental setup, each policy aimed to reduce the most uncertainty for a fixed number of test points. In reality, test-points are flown until mission resources, in the form of time or fuel, are exhausted. Thus, while the \fixedpolicy test-card policy exhibited a greater uncertainty reduction, the more relevant metric is the fuel normalized uncertainty reduction, where both adaptive policies significantly outperform the \fixedpolicy baseline. 

Overall, the results indicate a trade between absolute uncertainty reduction and fuel efficiency in the test-point constrained experiment: \fixedpolicy tends to achieve higher total reduction, while adaptive strategies achieve significantly higher uncertainty reduction per unit fuel. Within adaptive methods, \mctsshort significantly outperforms \greedyshort for both total uncertainty reduction and fuel efficiency.

\section{Conclusion and Future Work}
\label{sec:conclusion}

 We introduced a GP serving as a belief-state and reduced predictive uncertainty for a discretized set of flight conditions accounting for resource costs in the planner objective function. This study shows that posterior-aware adaptive maneuver selection improves fuel-normalized data efficiency in GP-based aerodynamic modeling under a fixed sortie budget while a fixed test-card can still deliver larger absolute uncertainty reduction for the same number of maneuvers. Sortie-summary statistics show statistically significant directional differences between adaptive policies and fixed test-cards in uncertainty reduction per unit fuel, suggesting that shallow lookahead primarily improves efficiency rather than absolute reduction. Among adaptive policies, \mctswivr is significantly more fuel efficient and yields greater uncertainty reduction than \greedywivr .

Future work will focus on the following: 
\begin{itemize}
    \item Relaxing the discrete maneuver assumption by embedding the planner in a continuous optimization framework that generates control time histories targeting high-uncertainty regions, enabling joint optimization of trajectory feasibility and information gain.
    
    \item Improving scalability, as GP updates become intractable with increasing data and dimensionality, by leveraging sparse GP methods and approximate planning strategies to reduce computational cost while preserving decision quality.
    
    \item Enhancing robustness to model misspecification through adaptive kernels, non-stationary models, or ensembles, along with improved uncertainty calibration to ensure reliable performance.

    \item Deepening multi-step lookahead to determine further improvement above and beyond that of the shallow lookahead studied here. 
\end{itemize}

\bibliographystyle{IEEEtran}
\bibliography{references}
\appendix

\section{Least Squares Approximation}

\subsection{Lift Model}
\label{app:lift}
This section summarizes the construction of the aerodynamic quantities and least-squares lift model used to infer angle of attack and related state variables from X-Plane test data. While the action space is capped at $M = 1.0$, the supersonic form is retained for completeness.

The lift coefficient is then computed directly from the force balance using flight data,
\begin{equation}
C_L(t)
=
\frac{m\, G(t)\, g}{\tfrac{1}{2}\rho V(t)^2 S},
\end{equation}
where $m$ is the aircraft mass, $\rho$ is air density, $V(t)$ is true airspeed at time $t$, and $S$ is the wing reference area.

Angle of attack is obtained by inverting a parametric least-squares lift model of the form
\begin{equation}
C_L(\alpha, M)
=
C_{L0}
+
a_0 \frac{\alpha}{\sqrt{1 - M^2}}
+
b_0 \frac{\alpha}{\sqrt{M^2 - 1}},
\end{equation}
where $C_{L0}$ is the zero-lift coefficient, $a_0$ is the subsonic lift-curve slope parameter, and $b_0$ is the supersonic lift-curve slope parameter. Solving this expression for $\alpha$ yields the regime-dependent inversion

\begin{equation}
\alpha(t)
=
\frac{C_L(t) - C_{L0}}{a_0 \sqrt{1 - M(t)^2}},
\quad M < 1,
\end{equation}
and
\begin{equation}
\alpha(t)
=
\frac{C_L(t) - C_{L0}}{b_0 \sqrt{M(t)^2 - 1}},
\quad M > 1.
\end{equation}

The parameters $\{C_{L0}, a_0, b_0\}$ are estimated using ordinary least squares by minimizing the squared error between measured and modeled lift coefficients,
\begin{equation}
\min_{C_{L0}, a_0, b_0}
\sum_{i=1}^{N}
\left(
C_L^{\mathrm{data}}(i)
-
C_L(\alpha_i, M_i)
\right)^2.
\end{equation}

To avoid the singularity at $M=1$, a smooth blending function is introduced,
\begin{equation}
w(M)
=
\frac{1}{2}\left[
1 + \tanh\left(\frac{M - M_t}{\Delta M}\right)
\right],
\end{equation}
with $M_t = 1.0$ and transition width $\Delta M= 0.08$. The final blended lift model is therefore written as
\begin{equation}
\begin{aligned}
C_L(\alpha, M)
&=
C_{L0}
+
(1 - w(M))\, a_0 \frac{\alpha}{\sqrt{1 - M^2}} \\
&\quad +
w(M)\, b_0 \frac{\alpha}{\sqrt{M^2 - 1}}.
\end{aligned}
\end{equation}

\subsection{Fuel Flow Model}
\label{app:fuel}
The fuel flow model is formulated as a log-linear regression in order to enforce positivity of the prediction and improve numerical conditioning across wide operating regimes. Specifically, the logarithm of the measured fuel flow rate is modeled as an affine function of Mach number and scaled altitude,
\begin{equation}
\ln(\dot{m}) = \beta_1 + \beta_2 M + \beta_3 \hat{h},
\end{equation}
where $M$ is the Mach number and $\hat{h} = h/10{,}000$ is altitude expressed in units of 10,000 ft to improve numerical conditioning.

The parameter vector $\beta = [\beta_1, \beta_2, \beta_3]^T$ is obtained using ordinary least squares on the log-transformed data,
\begin{equation}
\beta = X \backslash \ln(\dot{m}),
\end{equation}
where $X$ denotes the design matrix constructed from the regression features $[1, M, \hat{h}]$.

The resulting fuel flow prediction is recovered by exponentiating the linear model,
\begin{equation}
\dot{m}_{\mathrm{fuel}} = \exp\!\left(\beta_1 + \beta_2 M + \beta_3 \hat{h}\right),
\end{equation}
which guarantees $\dot{m}_{\mathrm{fuel}} > 0$ for all operating conditions.

During action evaluation, the maneuver fuel cost is computed by integrating the instantaneous fuel flow over the maneuver duration.

\end{document}

%% file: figures/traditional_vs_pointless2.tex
\begin{figure}[h]
\centering
\begin{tikzpicture}[
    scale=0.8,
    transform shape,
    font=\footnotesize,
    >=Stealth,
    point/.style={
        circle, draw=blue!60!black, thick,
        minimum size=10pt, inner sep=0pt
    },
    tolbox/.style={
        draw=blue!40, dashed, thin,
        minimum size=18pt, inner sep=0pt
    },
    statbox/.style={
        draw=gray!40, fill=gray!8,
        rounded corners=3pt,
        inner sep=2pt, align=center
    },
]

\begin{scope}

  \node[font=\small\bfseries] at (3.2, 6.0) {Traditional flight test};
  \node[text=gray!80, font=\footnotesize] at (3.2, 5.6)
    {fixed test points, rigid tolerances};

  \draw[draw=gray!50, rounded corners=3pt]
    (0.7, 0.8) rectangle (5.9, 5.2);

  \node[rotate=90, font=\footnotesize] at (-0.6, 3.0) {Mach};
  \node[font=\scriptsize, anchor=east] at (0.6, 4.8) {hi};
  \node[font=\scriptsize, anchor=east] at (0.6, 3.0) {med};
  \node[font=\scriptsize, anchor=east] at (0.6, 1.2) {lo};
  \node[font=\scriptsize, anchor=north] at (1.7, 0.75) {low};
  \node[font=\scriptsize, anchor=north] at (3.2, 0.75) {mid};
  \node[font=\scriptsize, anchor=north] at (4.9, 0.75) {high};
  \node[font=\footnotesize] at (3.2, 0.2) {angle of attack};

  \foreach \cx/\cy in {
      1.7/4.8, 3.2/4.8, 4.9/4.8,
      1.7/3.0, 3.2/3.0, 4.9/3.0,
      1.7/1.2, 3.2/1.2, 4.9/1.2}{
    \node[tolbox]  at (\cx,\cy) {};
    \node[point]   at (\cx,\cy) {};
  }

  \foreach \cx/\cy in {1.7/4.8, 3.2/4.8, 4.9/3.0, 1.7/1.2, 3.2/1.2}{
    \draw[red!70!black, thick]
      (\cx+0.18, \cy+0.18) -- (\cx+0.38, \cy+0.38);
    \draw[red!70!black, thick]
      (\cx+0.18, \cy+0.38) -- (\cx+0.38, \cy+0.18);
  }

  \foreach \cx/\cy in {4.9/4.8, 1.7/3.0, 3.2/3.0, 4.9/1.2}{
    \node[text=green!50!black, font=\scriptsize\bfseries]
      at (\cx+0.28, \cy-0.28) {\checkmark};
  }

  \node[statbox, text width=4.6cm] at (3.2, -0.6) {
    only 16\% of maneuvers are usable.\\
    avg.\ 6 repeats per test point.
  };

\coordinate (legTL) at (0.8, -1.4);

\node[point, scale=0.7] at (legTL) {};
\node[anchor=west, font=\scriptsize] at ($(legTL)+(0.35,0)$) {planned test point};

\draw[red!70!black, thick] 
  ($(legTL)+(0,-0.35)+(-0.12,0.12)$) -- ($(legTL)+(0,-0.35)+(0.12,-0.12)$);
\draw[red!70!black, thick] 
  ($(legTL)+(0,-0.35)+(-0.12,-0.12)$) -- ($(legTL)+(0,-0.35)+(0.12,0.12)$);
\node[anchor=west, font=\scriptsize] at ($(legTL)+(0.35,-0.35)$) {rejected};

\node[text=green!50!black, font=\scriptsize\bfseries] 
  at ($(legTL)+(0,-0.70)$) {\checkmark};
\node[anchor=west, font=\scriptsize] 
  at ($(legTL)+(0.35,-0.70)$) {accepted};

\end{scope}

\begin{scope}[yshift=-8.8cm]

  \node[font=\small\bfseries] at (3.2, 6.0) {Optimized approach};
  \node[text=gray!80, font=\footnotesize] at (3.2, 5.6)
    {arbitrary trajectories, all data used};

  \draw[draw=gray!50, rounded corners=3pt]
    (0.7, 0.8) rectangle (5.9, 5.2);

  \node[rotate=90, font=\footnotesize] at (-0.6, 3.0) {Mach};
  \node[font=\scriptsize, anchor=east] at (0.6, 4.8) {hi};
  \node[font=\scriptsize, anchor=east] at (0.6, 3.0) {med};
  \node[font=\scriptsize, anchor=east] at (0.6, 1.2) {lo};
  \node[font=\scriptsize, anchor=north] at (1.7, 0.75) {low};
  \node[font=\scriptsize, anchor=north] at (3.2, 0.75) {mid};
  \node[font=\scriptsize, anchor=north] at (4.9, 0.75) {high};
  \node[font=\footnotesize] at (3.2, 0.2) {angle of attack};

  \draw[teal!80!black, thick]
    (0.9, 4.4) .. controls (1.7, 3.6) and (2.4, 2.5) ..
    (3.0, 1.8) .. controls (3.8, 1.2) and (4.5, 1.3) .. (5.7, 0.95);
  \foreach \p in {(0.9,4.4),(1.5,3.8),(2.2,2.9),(2.8,2.0),
      (3.5,1.4),(4.3,1.25),(5.2,1.0),(5.7,0.95)}{
    \fill[teal!80!black] \p circle (2pt);
  }

  \draw[teal!55!black, thick]
    (0.9, 3.2) .. controls (1.8, 3.9) and (2.6, 3.4) ..
    (3.3, 3.0) .. controls (4.1, 2.6) and (4.8, 2.8) .. (5.7, 2.4);
  \foreach \p in {(0.9,3.2),(1.6,3.7),(2.4,3.5),(3.0,3.1),
      (3.7,2.8),(4.4,2.75),(5.1,2.5),(5.7,2.4)}{
    \fill[teal!55!black] \p circle (2pt);
  }

  \draw[teal!35!black, thick]
    (0.9, 4.9) .. controls (1.7, 4.6) and (2.5, 4.2) ..
    (3.1, 3.8) .. controls (3.9, 3.5) and (4.6, 3.7) .. (5.7, 3.5);
  \foreach \p in {(0.9,4.9),(1.6,4.65),(2.3,4.35),(2.8,4.0),
      (3.5,3.7),(4.2,3.65),(4.9,3.55),(5.7,3.5)}{
    \fill[teal!35!black] \p circle (2pt);
  }

  \node[statbox, text width=4.6cm] at (3.2, -0.6) {
    all data contributes to the GP\\
    aerodynamic model
  };

  \draw[teal!80!black, thick] (0.6, -1.5) -- (1.1, -1.5);
  \fill[teal!80!black] (0.85, -1.5) circle (2pt);
  \node[anchor=west, font=\scriptsize] at (1.2, -1.5) {trajectory};

  \fill[teal!60!black] (0.85, -1.8) circle (2.5pt);
  \node[anchor=west, font=\scriptsize] at (1.2, -1.8) {data point};

\end{scope}

\path[use as bounding box] (0.5, -10.0) rectangle (6.1, 6.2);

\end{tikzpicture}

\caption{Traditional flight testing uses discrete test points and rejects off-nominal data, whereas the optimized approach uses all data collected along arbitrary trajectories.
}
\label{fig:traditional_vs_pointless}
\end{figure}

%% file: references.bib
@String { icra        = {IEEE International Conference on Robotics and Automation (ICRA)} }

@String { ieeeciaig   = {IEEE Transactions on Computational Intelligence and AI in Games} }

@String { iros        = {IEEE/RSJ International Conference on Intelligent Robots and Systems (IROS)} }

@misc{harp2024pointless,
   title         = {A Data-Based Architecture for Flight Test without Test Points},
   author        = {D. Isaiah Harp and Joshua Ott and John Alora and Dylan Asmar},
   year          = {2025},
   eprint        = {2506.02315},
   archivePrefix = {arXiv},
   primaryClass  = {cs.LG},
   url           = {https://arxiv.org/abs/2506.02315},
}

@Article{Nelder1965,
  author    = {Nelder, John A. and Mead, Roger},
  title     = {A simplex method for function minimization},
  number    = {4},
  pages     = {308--313},
  volume    = {7},
  journal   = {The Computer Journal},
  publisher = {British Computer Society},
  year      = {1965},
}

@article{harp2025pigp,
   author  = {Isaiah Harp, D. and Ott, Joshua and Asmar, Dylan M. and Alora, John
              and Kochenderfer, Mykel J.},
   title   = {Physics-Informed {G}aussian Processes for Efficient Envelope Expansion},
   journal = {Journal of Aerospace Information Systems},
   year    = {2025},
   note    = {Articles in Advance},
   doi     = {10.2514/1.I011606},
}

@book{kochenderfer2022adm,
   author    = {Mykel J. Kochenderfer and Tim A. Wheeler and Kyle H. Wray},
   city      = {Cambridge},
   publisher = {MIT Press},
   title     = {Algorithms for Decision Making},
   year      = {2022},
}

@book{kochenderfer2019,
   author    = {Mykel J. Kochenderfer and Tim A. Wheeler},
   publisher = {MIT Press},
   title     = {Algorithms for Optimization},
   year      = {2019},
}

@book{rasmussen2006gp,
   author    = {Rasmussen, Carl Edward and Williams, Christopher K. I.},
   title     = {Gaussian Processes for Machine Learning},
   publisher = {MIT Press},
   year      = {2006},
}

@article{browne2012mcts,
  title={A survey of monte carlo tree search methods},
  author={Browne, Cameron B and Powley, Edward and Whitehouse, Daniel and Lucas, Simon M and Cowling, Peter I and Rohlfshagen, Philipp and Tavener, Stephen and Perez, Diego and Samothrakis, Spyridon and Colton, Simon},
  journal=ieeeciaig,
  volume={4},
  number={1},
  pages={1--43},
  year={2012},
  publisher={IEEE}
}

@article{settles2009active,
   author  = {Settles, Burr},
   title   = {Active Learning Literature Survey},
   journal = {Computer Sciences Technical Report 1648, University of
              Wisconsin--Madison},
   year    = {2009},
}

@article{paninski2005infodesign,
   author  = {Paninski, Liam},
   title   = {Asymptotic Theory of Information-Theoretic Experimental Design},
   journal = {Neural Computation},
   volume  = {17},
   number  = {7},
   pages   = {1480--1507},
   year    = {2005},
}

@article{silver2010pomcp,
   author  = {Silver, David and Veness, Joel},
   title   = {{Monte-Carlo} Planning in Large {POMDPs}},
   journal = {Advances in Neural Information Processing Systems},
   volume  = {23},
   year    = {2010},
}

@article{ross2008online,
   author  = {Ross, St\'{e}phane and Pineau, Joelle and Paquet, S\'{e}bastien
              and Chaib-draa, Brahim},
   title   = {Online Planning Algorithms for {POMDPs}},
   journal = {Journal of Artificial Intelligence Research},
   volume  = {32},
   pages   = {663--704},
   year    = {2008},
}

@book{phillips2009mechanics,
  author    = {Phillips, Warren F.},
  title     = {Mechanics of Flight},
  edition   = {2nd},
  publisher = {Wiley},
  address   = {Hoboken, NJ},
  year      = {2009}
}

@article{grauer2015morelli,
   author  = {Grauer, Jared A. and Morelli, Eugene A.},
   title   = {Generic Global Aerodynamic Model for Aircraft},
   journal = {Journal of Aircraft},
   volume  = {52},
   number  = {1},
   pages   = {13--20},
   year    = {2015},
   doi     = {10.2514/1.C032615},
}

@article{karaman2011rrt,
   author  = {Karaman, Sertac and Frazzoli, Emilio},
   title   = {Sampling-Based Algorithms for Optimal Motion Planning},
   journal = {International Journal of Robotics Research},
   volume  = {30},
   number  = {7},
   pages   = {846--894},
   year    = {2011},
   doi     = {10.1177/0278364911406761},
}

@article{vanlier2012bayesian,
   author  = {Vanlier, Joep and Tiemann, Christian A. and Hilbers, Peter A. J.
              and van Riel, Natal A. W.},
   title   = {A {Bayesian} Approach to Targeted Experiment Design},
   journal = {Bioinformatics},
   volume  = {28},
   number  = {8},
   pages   = {1136--1142},
   year    = {2012},
   doi     = {10.1093/bioinformatics/bts092},
}

@inproceedings{webb2013kinodynamic,
   author    = {Webb, Dustin J. and van den Berg, Jur},
   title     = {Kinodynamic {RRT*}: Asymptotically Optimal Motion Planning for
                Robots with Linear Dynamics},
   booktitle = {IEEE International Conference on Robotics and Automation (ICRA)},
   pages     = {5054--5061},
   year      = {2013},
   doi       = {10.1109/ICRA.2013.6631299},
}

@inproceedings{delmas2019mcts,
   author    = {Delmas, R\'{e}mi and Loquen, Thomas and Boada-Bauxell, Josep
                and Carton, Mathieu},
   title     = {An Evaluation of {Monte-Carlo} Tree Search for Property Falsification
                on Hybrid Flight Control Laws},
   booktitle = {Numerical Software Verification (NSV)},
   series    = {Lecture Notes in Computer Science},
   volume    = {11652},
   pages     = {45--59},
   publisher = {Springer},
   year      = {2019},
   doi       = {10.1007/978-3-030-28423-7_3},
}

@article{riviere2024montecarlo,
  author  = {Rivi{\`e}re, Benjamin and Lathrop, John and Chung, Soon-Jo},
  title   = {Monte {C}arlo tree search with spectral expansion for planning with dynamical systems},
  journal = {Science Robotics},
  volume  = {9},
  number  = {97},
  pages   = {eado1010},
  year    = {2024},
  doi     = {10.1126/scirobotics.ado1010}
}

@inproceedings{guo2022auv,
  title={Informative path planning for auv-based underwater terrain exploration with a {POMDP}},
  author={Zhang, Shi and Cui, Rongxin and Yan, Weisheng and Li, Yinglin},
  booktitle={China Automation Congress (CAC)},
  pages={4756--4761},
  year={2021},
  organization={IEEE}
}

@article{ott2024ipp,
   author  = {Ott, Joshua and Kochenderfer, Mykel J. and Boyd, Stephen},
   title   = {Approximate Sequential Optimization for Informative Path Planning},
   journal = {Robotics and Autonomous Systems},
   volume  = {182},
   pages   = {104814},
   year    = {2024},
   doi     = {10.1016/j.robot.2024.104814},
}

@inproceedings{gammell2014informed,
  title={Informed RRT*: Optimal sampling-based path planning focused via direct sampling of an admissible ellipsoidal heuristic},
  author={Gammell, Jonathan D. and Srinivasa, Siddhartha S. and Barfoot, Timothy D.},
  booktitle={IEEE/RSJ International Conference on Intelligent Robots and Systems (IROS)},
  pages={2997--3004},
  year={2014},
  organization={IEEE}
}

@inproceedings{popovic2020informative,
  author    = {Popovi\'c, Marija and Vidal-Calleja, Teresa and Chung, Jen Jen and Nieto, Juan and Siegwart, Roland},
  title     = {Informative path planning for active field mapping under localization uncertainty},
  booktitle = {Proc. IEEE Int. Conf. Robotics and Automation (ICRA)},
  pages     = {10751--10757},
  year      = {2020}
}

@misc{akemoto2025informativepathplanningexplore,
      title={Informative Path Planning to Explore and Map Unknown Planetary Surfaces with Gaussian Processes}, 
      author={Ashten Akemoto and Frances Zhu},
      year={2025},
      eprint={2503.16613},
      archivePrefix={arXiv},
      primaryClass={cs.RO},
      url={https://arxiv.org/abs/2503.16613}, 
}
